\documentclass[twocolumn,prb,showpacs,superscriptaddress,floatfix,nofootinbib]{revtex4}

\usepackage{amssymb}
\usepackage{amsmath}
\usepackage{graphicx}
\usepackage{dcolumn}
\usepackage{bm}
\usepackage{color}

\newcommand{\units}[1]{\ensuremath{\,\mathrm{#1}}}

\newcommand{\fracp}[3][]{\displaystyle\frac{\partial^{#1}#2}{\partial#3^{#1}}}

\begin{document}

\title{Method for reliable realization of a $\varphi$ Josephson junction}

\author{N.~G.~Pugach}
\email{pugach@magn.ru}
\affiliation{Faculty of Physics, M.V. Lomonosov Moscow State University, 119992 Leninskie Gory, Moscow, Russia}

\author{E.~Goldobin}
\author{R.~Kleiner}
\author{D.~Koelle}
\affiliation{%
  Physikalisches Institut--Experimentalphysik II and Center for Collective Quantum Phenomena,
  Universit\"at T\"ubingen,
  Auf der Morgenstelle 14,
  D-72076 T\"ubingen, Germany
}

\date{\today}

\begin{abstract}

  We propose a method to realize a $\varphi$ Josephson junction by combining alternating 0 and $\pi$ parts (sub junctions) with an intrinsically non-sinusoidal current-phase relation (CPR). Conditions for the realization of the $\varphi$ ground state are analyzed. It is shown that taking into account the non-sinusoidal CPR for a ``clean'' junction with a ferromagnetic (F) barrier, one can significantly enlarge the domain (regime of suitable F-layer thicknesses) of the $\varphi$ ground state and make the practical realization of $\varphi$ Josephson junctions feasible. Such junctions may also have two different stable solutions, such as 0 and $\pi $, 0 and $\varphi $, or $\varphi $ and $\pi $.

\end{abstract}

\maketitle

\section{Introduction}
\label{sec:Int}

The interest on Josephson junction (JJ) devices with a ferromagnetic (F) barrier has been continuously increasing during the last years \cite{Buzdin:2005:Review:SF,KuprReiew}. The critical current density $j_c$ of a SFS (S -- superconductor)  junction exhibits damped oscillations as a function of the F-layer thickness $d_F$ so that the Josephson phase $\phi $ can be 0 or $\pi$ in the ground state \cite{Buzdin:2005:Review:SF,KuprReiew}. $\pi$ junctions can be used as (non-dischargeable) on-chip $\pi$-phase batteries for self-biasing various electronic circuits in the classical and quantum domains, e.g.~self-biased RSFQ logic\cite{Ortlepp:2006:RSFQ-0-pi} or flux qubits\cite{Ioffe:1999:sds-waveQubit,OurQubit}. In addition, for quantum circuits self-biasing also decouples the circuit from the environment and improves decoherence figures, e.g., as in the quiet qubit.\cite{Ioffe:1999:sds-waveQubit,Yamashita:2005:pi-qubit:SFS+SIS,Yamashita:2006:pi-qubit:3JJ}
In classical circuits a phase battery may also substitute the conventional inductance and substantially reduce the size of an elementary cell\cite{Ustinov:2003:RSFQ+pi-shifters}. Some of these proposals were already realized practically\cite{Ortlepp:2006:RSFQ-0-pi,RyazanovPiCirc2009}.

In this context it is even more interesting to create a Josephson junction with an arbitrary phase difference $\varphi$ ($0<|\varphi|<\pi$) in the ground state --- a so-called $\varphi$ junction\cite{BuzdinKoshelev}. In addition to providing an arbitrary phase bias, such $\varphi$ junctions  have rather interesting physical properties such as two critical currents, non-Fraunhofer Josephson current dependence on an external magnetic field, half-integer Shapiro steps and an unusual behavior when embedded in a SQUID loop. In \emph{long} $\varphi$ junctions two types of mobile Josephson vortices carrying fractional magnetic flux $\Phi_1<\Phi_0$ and $\Phi_2=\Phi_0-\Phi_1$ ($\Phi _{0}\approx2.07\times10^{-15}\units{Wb}$ is the magnetic flux quantum) may exist, resulting in  half integer zero field steps, two critical values of magnetic field penetration and other unusual properties.\cite{Goldobin2harm}

To obtain a $\varphi$ junction the Josephson energy density
\begin{equation}
  E_{J}=\frac{\Phi _{0}}{2\pi }\int\limits_{0}^{\phi }j(\phi ^{\prime})d\phi ^{\prime }
  \label{Ej}
\end{equation}
should have a local minimum at $\phi=\varphi$ ($0<|\varphi|<\pi$). To achieve this, the current-phase relation (CPR) $j(\phi)$ should be different from the usual sinusoidal one
\begin{equation}
  j(\phi)=j_{1}\sin \phi
  , \label{sinCPR}
\end{equation}
where $j$ is the Josephson (super)current density and $j_1$ is the critical current density. It has been shown that a $\varphi$ junction can be realized with a second harmonic in the CPR, i.e.,
\begin{equation}
  j(\phi )=j_{1}\sin \phi +j_{2}\sin 2\phi
  , \label{CPR2}
\end{equation}%
if $j_{2}<0$ and\cite{Goldobin2harm}
\begin{equation}
  \left\vert \frac{2j_{2}}{j_{1}}\right\vert >1
  . \label{condition2}
\end{equation}
Note that in Eq.~\eqref{CPR2} $j_1$ is not anymore the critical current density, but just the amplitude of the first harmonic and $j_2$ is the amplitude of the second harmonic. The critical current density $j_c$ is then determined by local maxima of $j(\phi)$ as shown below.

A non-sinusoidal CPR is not so exotic for Josephson junctions with a ferromagnetic barrier\cite{KuprReiew}, as  demonstrated recently in several experiments\cite{Ryazanov2harmNovgorod,Sellier:2004:SFS:HalfIntShapiro}. For example, in the simplest case of SFS junctions consisting of pure S and F metals, the CPR is strongly non-sinusoidal at a temperature $T \ll T_c$, where $T_c$ is the critical temperature of the S metal. Different scattering mechanisms in the F-layer also influence the CPR: usual (non-pair breaking) scattering (``dirty'' limit)\cite{Buzdin:1982,Buzdin2-3D}; spin-flip scattering\cite{Buzdin:2005:0-pi-trans} and scattering of electrons from the $s$ to the $d$-band\cite{OurSFS,OurQubit}. In addition, the transparency of the interfaces or the presence of an extra insulating (I) dielectric layer, like in SIFS or SIFIS junctions, influences the CPR as well\cite{KuprCPR,KuprFominovCPR}. However, the theoretical models\cite{KuprReiew,Buzdin2-3D,Buzdin:2005:0-pi-trans,KuprCPR,OurSFS,OurQubit} that take into account different scattering mechanisms give similar results: The amplitude of the second harmonic oscillates and decreases with $d_F$ twice faster than the first one. To satisfy (\ref{condition2}) one is tempted to choose $d_F$ in the vicinity of 0-$\pi$ transition, where $|j_{1}|\ll |j_{2}|$. Unfortunately, there $j_{2}>0$\cite{Buzdin2-3D,Buzdin:2005:0-pi-trans,OurSFS,OurQubit}, which excludes a $\varphi$ ground state.

A technique to create a negative second harmonics artificially was proposed recently\cite{Mints2harm,BuzdinKoshelev}. By using one of the available 0-$\pi$ junction technologies\cite{Weides:0-piLJJ,Smilde:ZigzagPRL}, one fabricates a Josephson junction with alternating short 0 ($j_1>0$) and $\pi$ ($j_1<0$) regions, which all have the simple CPR \eqref{sinCPR}. Here, an {\em effective} second harmonic with negative amplitude is generated for the averaged phase, which is slowly varying on the scale of the facet length\cite{Mints2harm,BuzdinKoshelev}. Let the alternating 0 and $\pi$ regions have the facet lengths $a$ and $b$, respectively, so that $a\approx b$. It is assumed that $a,b<\lambda_J^a,\lambda_J^b$, where $\lambda _{J}^a=\lambda _{J}(j_1^a)$ and $\lambda _{J}^b=\lambda _{J}(j_c^b)$ are the Josephson penetration depth in the corresponding parts, with

\begin{equation}
  \lambda _{J}(j_c)=\sqrt{\frac{\Phi _{0}}{2\mu_0 \pi d' \left| j_c \right| } }
  . \label{lambdaJ}
\end{equation}
Here $\mu_0 d'$ is the effective inductance per square of the junction electrodes.

The effective second harmonic has a maximum amplitude if $a=b\approx \lambda _{J}$, and if the values of the critical current densities in corresponding regions $j_{1}^a\approx -j_{1}^b$, see Ref.~\onlinecite{BuzdinKoshelev,Goldobin2harm}. The effective second harmonic amplitude is large enough only if $j_1^a$ and $j_1^b$ are very close by absolute values, demanding to choose thicknesses $d_{F,a}$ and $d_{F,b}$ with very high precision less than 1\,\AA\ for usual ``dirty'' SFS junctions. This precision is not achievable technologically; therefore, a controllable $\varphi $-junction is quite hard to realize in this way.

However, ferromagnetic Josephson junctions already have some intrinsic second harmonic. Thus, the question is what will be the effective CPR in multifacet junctions if one takes into account such an intrinsic second harmonic.
Usually, $j_{2}<0$ in the range of $d_F$ where $j_{1}$ is rather large. The idea of the method used here is the following. We use two thicknesses $d_{F,a}$ and $d_{F,b}$ where $j_{2}(d_{F,a})<0$ and $j_{1}(d_{F,a})\approx-j_{1}(d_{F,b})$. By making a step-like F-layer changing between $d_{F,a}$ and $d_{F,b}$, we effectively cancel the large first harmonic and create an effective negative second harmonic which adds up with intrinsic second harmonic.

The paper is organized as follows. In Sec.~\ref{sec:Model} we describe the model and derive the equations for the effective phase using an averaging procedure over the rapid oscillations of the phase on the length scales set by $a$ and $b$ . Section \ref{sec:Discuss} contains the discussion and results of calculations done for a ``clean'' SFS junction. Section \ref{sec:Con} concludes this work.

\section{Model}
\label{sec:Model}

Mints and coauthors\cite{Mints2harm,Mints:2000:SelfGenFlux@GB,Mints:2001:FracVortices@GB} considered 1-dimensional junctions (along $x$ direction) with a critical current density $j_{1}(x)$ which is a random or periodic function changing on a length-scale $a,\,b\ll \lambda _{J}^{a,b}=\lambda _{J}(j_{1a,b})$. Buzdin and Koshelev  \cite{BuzdinKoshelev,Goldobin2harm} have found an exact solution of this problem if $j_{1}(x)$ in \eqref{sinCPR} alternates between the constant values $j_{1}^a$ and $j_{1}^b$. Both groups assumed that $j_{2}=0$ within each region. Instead, here we assume that the CPR $j(d_F(x),\phi)$ is non-sinusoidal as a function of $\phi$ within each region, and alternates between $j(d_{F,a},\phi)$ and $j(d_{F,b},\phi)$ as a function of $x$. Then the problem of calculating the phase distribution along such a non-uniform structure can be reduced to the well-known problem of a non-linear oscillator behavior \cite{LLmechanics}.

The Josephson current in every region, as an odd function of
the phase, can be expanded in a series of harmonics
\begin{equation}
  j(d_F(x),\phi )=\sum_{n=1}^{\infty }j_{n}(d_F(x))\sin n\phi
  ,  \label{CPR}
\end{equation}
and the problem is solved in a general form. Here $j_{n}$ denotes the amplitude of the \emph{intrinsic} $n$-th harmonic of the current density.

We assume $\xi _{F}\ll a,b\leq \lambda_{J}$, were, $\xi _{F}$ is the ferromagnetic coherence length, i.e.~the characteristic length for the critical current density
nonuniformity at the region of the step-like change of junction properties \cite{ourJETPL}.
Let us consider one period of the structure $[-a;b)$ with F-layer thickness
\begin{equation}
    d_F(x)=\left\{
    \begin{array}{ll}
      d_{F,a},  &  x\in[-a,0)\\
      d_{F,b},  &  x\in[0,b)\quad . \\
    \end{array}
  \right.
\end{equation}
We rewrite \eqref{CPR} to separate the average and the oscillating parts of the Josephson current, i.e.,
\begin{equation}
  j(x,\phi )= \sum_{n=1}^{\infty }
    \left\langle j_{n}\right\rangle [1+g_{n}(x)]\sin n\phi
  , \label{SLcurrent}
\end{equation}
where $j(x,\phi)=j(d_F(x),\phi)$.
The average value of any function $f(x)$ is defined as
\begin{equation}
  \left\langle f\right\rangle =\frac{1}{a+b}\int\limits_{-a}^{b}f(x)dx.
\end{equation}
Then the averaged supercurrent of the $n$-th harmonic is
\begin{equation}
  \langle j_{n}\rangle =\frac{aj_{n}(d_{F,a})+b j_n(d_{F,b})}{a+b}
  , \label{Eq:j_n:av}
\end{equation}
and the corresponding oscillating part is
\begin{equation}
  g_{n}(x) = (j_{n}(x)-\left\langle j_{n}\right\rangle )/\left\langle j_{n} \right\rangle
  , \label{Eq:g_n}
\end{equation}
so that $\left\langle g_{n}\right\rangle =0$ by definition. Here $j_n(x,\phi)=j_n(d_F(x),\phi)$.

One can represent the Josephson phase $\phi(x)$ as a sum of a slow component $\psi(x)$, changing on a distance $\sim \Lambda _{J}=\lambda _{J}(\left\langle j_{1}\right\rangle )$, see Eq.~\eqref{lambdaJ}, and a rapid component $\zeta(x)$, changing on a distance $\sim a,b$ , i.e.,
\begin{equation}
  \phi (x)=\psi +\zeta (x)
  . \label{SLphase}
\end{equation}
Here we assume that the junction is short ($l\ll \Lambda_J$) and do not write the $x$ dependence for $\psi$. We also assume that the average of fast phase oscillations is vanishing, i.e.,
\begin{equation}
  \left\langle \zeta \right\rangle =0,  \label{rapidPh-small}
\end{equation}
and their amplitude is small
\begin{equation}
  \left\langle \left\vert \zeta \right\vert \right\rangle \ll 1.
  \label{rapidPh-small2}
\end{equation}

The ground state of the Josephson junction is determined by the Ferrel-Prange equation
\begin{equation}
  \Lambda _{J}^{2}\frac{\partial ^{2}\phi }{\partial x^{2}}
  =\frac{j(x,\phi )}{\left\vert \left\langle j_{1}\right\rangle \right\vert }.
  \label{FP-Eq}
\end{equation}
Substituting Eqs.~(\ref{SLcurrent}) and (\ref{SLphase}) into Eq. (\ref{FP-Eq})
and keeping the terms up to first order in $\zeta (x)$\ we can obtain
equations for the rapid phase $\zeta$:

\begin{equation}
  \Lambda _{J}^{2}\frac{\partial ^{2}\zeta }{\partial x^{2}}%
  -\sum_{n=1}^{\infty }\beta _{n}g_{n}(x)\sin n\psi =0.  \label{rapidPh}
\end{equation}
To obtain the equation for the slow phase $\psi$, we average Eq.~\eqref{FP-Eq} over the
length $(a+b)\ll \Lambda _{J}$ and get

\begin{equation}
  \Lambda _{J}^{2}\fracp[2]{\psi}{x}
  -\sum_{n=1}^{N}
    [\beta _{n}\sin n\psi +n\beta _{n}\left\langle g_{n}\zeta \right\rangle \cos n\psi] = 0
  ,  \label{slowPh}
\end{equation}
where $\beta _{n}=\langle j_{n}\rangle /|\langle j_{1} \rangle |$. The number of harmonics $N$, that is reasonable to take into account within the given approximation, follows from the condition $N\left\vert \zeta (x)\right\vert \ll 1$. We have to find the function $\zeta (x)$ from equation (\ref{rapidPh}), calculate average values $\left\langle g_{n}\zeta \right\rangle $, and substitute them into (\ref{slowPh}). Since for step-like $d_F(x)$ the Josephson current (\ref{CPR}) is a step-like function of $x$, Eq.~(\ref{rapidPh}) has the form $\partial ^{2}\zeta /\partial x^{2}=const$ on every interval $[-a;0)$ and $[0;b)$. Its solution is a parabolic segment. The function $\zeta (x)$ must be continuous at $x=0$, and at the edges, i.e., it must satisfy the boundary condition $\zeta (-a)=\zeta (b)$. Moreover, it should satisfy Eq. (\ref{rapidPh-small}).

It is convenient to expand the rapid phase $\zeta(x)$, as a solution of Eq. (\ref{rapidPh}), into a series
\begin{equation}
  \zeta (x)=\sum_{n=1}^{\infty }\zeta _{n}(x)\sin n\psi .
\end{equation}%
From this, the average values are calculated as
\begin{equation}
  \beta _{n}\left\langle g_{n}\zeta _{k}\right\rangle =-\frac{2\alpha \delta j_{n}\delta j_{k}}{\left\vert \left\langle j_{1}\right\rangle \right\vert }%
  ; \quad n,k=1,2...
  \label{averages}
\end{equation}
where $\delta j_n \equiv j_{n}(d_{F,a})-j_{n}(d_{F,b})$ and

\begin{equation}
\alpha \equiv \frac{a^{2}b^{2}}{24\Lambda _{J}^{2}(a+b)^{2}\left\vert \left\langle
j_{1}\right\rangle \right\vert }  \quad .\label{alpha}
\end{equation}
By definition (\ref{lambdaJ}), $\Lambda _{J}^{2}\left\vert \left\langle j_{1}\right\rangle
\right\vert =\lambda _{J}^{2}\left\vert j_{1a}\right\vert $,
where for brevity $j_{1a}\equiv j_{1}(d_{F,a}), \lambda _{J}\equiv\lambda _{J}(j_{1a})$.

The dependence of the Josephson current on the slow phase, that changes on a large distance of the order of $\Lambda _{J}$, follows from equation (\ref{slowPh}) as
\begin{equation}
J(\psi )=\left\vert \left\langle j_{1}\right\rangle \right\vert
\left[\sum\limits_{n}\beta _{n}\sin n\psi +\sum\limits_{n,k}n\beta
_{n}\left\langle g_{n}\zeta _{k}\right\rangle \cos n\psi \sin k\psi\right]
\label{J(ksi)}
\end{equation}%
The CPR (\ref{J(ksi)}) contains contributions of two types:
intrinsic harmonics, that are given by the first term of (\ref{J(ksi)}) and
the effectively generated ones, that are given by the second term. Intrinsic contributions are defined by the CPR (\ref{CPR}) of the junction regions, while generated
contributions effectively appear as a result of averaging over fast
oscillations.

The amplitudes of the effectively generated harmonics, which are proportional to average values (\ref{averages}), are largest by absolute value if $\alpha $ and $\left\vert \delta j_{n}\right\vert $\ reach their maximum. This happens if the lengths of $a$ and $b$ facets have the largest possible size which still allows averaging, i.e,
\begin{equation}
  a=b\approx \lambda _{J}  \label{a=b}
\end{equation}
and
\begin{equation}
  j_{n}(d_{F,a})\approx -j_{n}(d_{F,b})  \label{Ja=Jb}
\end{equation}
Condition (\ref{Ja=Jb}) ensures $\left\vert \left\langle
j_{1}\right\rangle \right\vert \ll \left\vert j_{1a}\right\vert $ and
consequently $\Lambda _{J}\gg \lambda _{J}$. It was shown
\cite{BuzdinKoshelev}, that
even if the equality (\ref{a=b}) holds exactly, condition (\ref{rapidPh-small2}) is satisfied. In this case $\alpha = 1/96\left\vert j_{1a}\right\vert $ [c.f.~(\ref{alpha})].

In all theoretical models developed up to now \cite{KuprReiew,Buzdin2-3D,Buzdin:2005:0-pi-trans,KuprCPR,OurSFS,OurQubit} the second harmonic was usually considered to be much smaller than the first one (except for the points of the 0-$\pi $ transition, where $j_{1}\rightarrow 0$), with an even smaller third harmonic. So, the expression (\ref{CPR}) can be considered as a Tailor expansion in some small parameter. Then, keeping terms of the same order of this small parameter, we obtain $n+k=N+1$ effective harmonics in the CPR of the multifacet 0-$\pi$ junction.

The number $N$ of harmonics, that is reasonable to take into
account, follows from the estimate of the short-range phase $N\max
\left\vert \zeta (x)\right\vert \ll 1$. As $\zeta (x)$ has a parabolic form, it takes its maximum absolute value at the center of every interval $%
-a/2$ and $b/2$. With the estimate
\begin{equation}
  \sum_{n=1}^{\infty }
  \beta_{n}g_{n}(x)\sin n\psi \sim g_{1}
  \label{estimate-zeta-max1}
\end{equation}
one obtains
\begin{equation}
  \left\vert \zeta (-a/2)\right\vert \sim
  \frac{ab(a+2b) |\delta j_{1}|}{24\Lambda_J^2 (a+b)|\langle j_{1} \rangle|}.
  \label{estimate-zeta-max2}
\end{equation}
This is maximized if $a=b=\lambda _{J}$. Since $\max \left\vert \delta
j_{1}\right\vert \approx2\left\vert j_{1a}\right\vert$, we estimate $\max \left\vert \zeta (x)\right\vert \approx 1/8$.
Thus, as a reasonable choice one can take $N=3$. So, within the above approximations, it is reasonable to consider three intrinsic harmonics with amplitudes $j_{1..3}$ and four generated harmonics, i.e.
\begin{equation}
  J(\psi )=\sum\limits_{n=1}^{4}J_{n}\sin n\psi
  \label{CPR4}
\end{equation}
with
\begin{eqnarray}
J_{1} &=&\left\langle j_{1}\right\rangle +\alpha \delta j_{1}\delta j_{2}
\nonumber \\
J_{2} &=&\left\langle j_{2}\right\rangle -\alpha \delta j_{1}^{2}+2\alpha
\delta j_{1}\delta j_{3}  \nonumber \\
J_{3} &=&\left\langle j_{3}\right\rangle -3\alpha \delta j_{1}\delta j_{2}
\label{J1-4} \\
J_{4} &=&-2\alpha \delta j_{2}^{2}-4\alpha \delta j_{1}\delta j_{3}\;.
\nonumber
\end{eqnarray}
Here $J_{n}$ denotes the total amplitude of the $n$-th harmonic in the CPR for the averaged phase.
These expressions reduce to the earlier results\cite{BuzdinKoshelev,Goldobin2harm} if one takes
into account only the intrinsic first harmonic.
The expression for every harmonic is a simple sum of the corresponding average intrinsic harmonic and the generated part.

\section{Results and Discussion}
\label{sec:Discuss}

We first address the question which type of ferromagnetic junction with 0-$\pi$ facets can satisfy the conditions for a $\varphi $ junction in the best way. The critical current density $j_1$ of a Josephson junction with a ferromagnetic barrier having a complex coherence length $\xi _{F}$ ($\xi_{F}^{-1}=\xi _{1}^{-1}+i\xi _{2}^{-1}$) decays $\propto\exp (-d_F/\xi_1)$ and oscillates with a period\cite{KuprReiew} $2\pi \xi_2$ in the ``dirty'' limit (electron mean free path $l\ll \xi _{1,2}$). The decay length $\xi_1$ depends on $l$ as well as on a pair-breaking scattering length in the ferromagnet (spin-flip scattering\cite{Buzdin:2005:0-pi-trans} or scattering into the $d$-band\cite{OurEuLett}; both influence the CPR). Pair-breaking scattering leads to \cite{Oboznov:2006:SFS-Ic(dF),OurSFS}  $\xi _{1}<\xi _{2}$ and the Josephson current decays so rapidly in the ``dirty'' limit, that there are only very tiny regions of $d_{F,a}$ and $d_{F,b}$ on the dependence $j_{1}(d_F)$ where $j_1(d_{F,a}) \approx -j_1(d_{F,b})$.
In the ``clean'' limit the Josephson current decays much more slowly with increasing $d_F$. The cleaner is the ferromagnet, the larger is $l$, and the slower is the decay; c.f.~Eqs.~(23) and (24) with Eq.~(7) from Ref.~\onlinecite{Buzdin2-3D}, or see the discussion in Ref.~\onlinecite{OurEuLett}. In the limit $l\gg (d_F,\xi_{1,2})$ the critical current density decreases as \cite{Buzdin:1982} $1/d_F$.
Moreover, a ``clean'' SFS junction has a non-sinusoidal CPR, and its second harmonics $j_2<0$ in some regions of $d_F$ far from 0-$\pi $ transitions, that can effectively help to satisfy the conditions for the realization of a $\varphi$ junction. It was shown in different models, that the second harmonic $j_2(d_F)$ decays and oscillates with $d_F$ twice faster than $j_1(d_F)$.
Therefore, it is reasonable to take an F-layer with $d_{F,a}$ and $d_{F,b}$ of the order of a few $\xi_2$. Usually, $\xi_2=v_{F}/2E_\mathrm{ex}$, where $v_{F}$
is the Fermi velocity and $E_\mathrm{ex}$ is the exchange magnetic energy in the ferromagnet\cite{Buzdin2-3D,OurEuLett,BergeretVolkovEfetovPRB2001}. It was established experimentally, that for pure ferromagnetic metals $\xi _{2}$ is largest for Ni \cite{RobinsonPRL97,RobinsonPRB76,Blum:2002:IcOscillations,WeidesRyazanov2009}.
Higher harmonics also decay rapidly with increasing
temperature\cite{Buzdin2-3D,Radovic:2001:0&pi-JJ}. We have also checked this statement for models described in \cite{Buzdin:1982,OurSFS}. It is clear that by approaching $T_{c}$ from below the superconducting gap $\Delta \rightarrow 0$, equations become linear and
their simple exponential solutions yield only a sinusoidal CPR. Therefore, the most promising strategy for realization of a $\varphi$ junctction is to use a pure SFS junction with a thin Ni layer at low temperature ($\lesssim 0.1 T_c$).
The model describing the Josephson effect in such junctions was established long ago\cite{Buzdin:1982}. It is based on the solution of the Eilenberger equations.

What is the measurable critical current density $J_C$ (maximum supercurrent) of a Josephson junction with a non-sinusoidal CPR? It is not anymore $|j_{1}|=|j(\pi /2)|$ as follows from (\ref{sinCPR}), but the local maximum of the expression (\ref{CPR}) with respect to $\phi$.
Note that (\ref{CPR}) may allow several local extrema in the interval $[0,2\pi)$.

\begin{figure}[b]
  \begin{center}
    \includegraphics{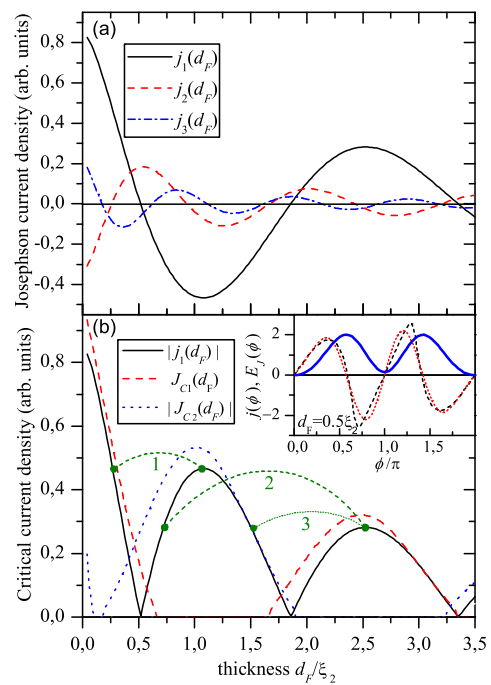}
  \end{center}
  \caption{(Color online)
    Properties of a uniform SFS Josephson junction in the ''clean`` limit at $T=0.1T_{c}$: (a) amplitudes of the first three harmonics $j_1,\,j_2,\,j_3$ of Eq.~(\ref{CPR}) as a function of the ferromagnet thickness $d_F$. (b) measurable critical current densities $J_{C1}$ and $|J_{C2}|$ vs $d_F$, which are realized at the phase $0<\phi<\pi$ and $\pi<\phi<2\pi$, respectively.
    Inset shows for $d_F=0.5\xi_{2}$ the CPR $j(\phi)$ from (3) in Ref.~\onlinecite{Buzdin:1982} (dashed line) and from the approximation by the first 3 harmonics (dotted line), with the corresponding $E_J (\phi)$ dependence (solid line).
    For comparison, in (b) the critical current $|j_1(d_F)|$ of a junction with only the first harmonic in the CPR is also shown. Here, pairs of points (connected by arcs) where $j_{1}(d_{F,a})= -j_{1}(d_{F,b})$  correspond to a $\varphi$ ground state of a multifacet SFS junction with alternating thicknesses $d_{F,a},\, d_{F,b}$  (see areas 1,2,3 in Fig.~\ref{fig:GroundStates}(a)).
  }
  \label{fig:j(d)}
\end{figure}

Figure \ref{fig:j(d)}(a) shows the three first harmonics vs F-layer thickness and Fig.~\ref{fig:j(d)}(b) shows the corresponding $J_C(d_F)$ for the model of an SFS junction as described in Ref.~\onlinecite{Buzdin:1982} (see also Ref.~\onlinecite{Buzdin2-3D}). It is interesting to note, that near a 0-$\pi $ transition, when the first harmonic is small, $j(\phi )$ may have two different local maxima, and the measurable $J_{C}(d_F)$ has two different values $J_{C1},\,|J_{C2}|$  depending on the initial state of the junction \cite{Goldobin2harm} (see Fig.~\ref{fig:j(d)}(b)).
An example of a CPR $j(\phi)$ with two different maxima and the corresponding Josephson energy $E_J(\phi)$ (\ref{Ej}) near a 0-$\pi$ transition are presented in Fig.~\ref{fig:j(d)}(b)(inset). Thus, even for a uniform Josephson junction (where the $\varphi $-ground state is impossible) close to a 0-$\pi $ transition, two different values of critical current density could be realized if its CPR differs enough from the sinusoidal one.

\begin{figure*}[!htb]
  \begin{center}
    {\includegraphics{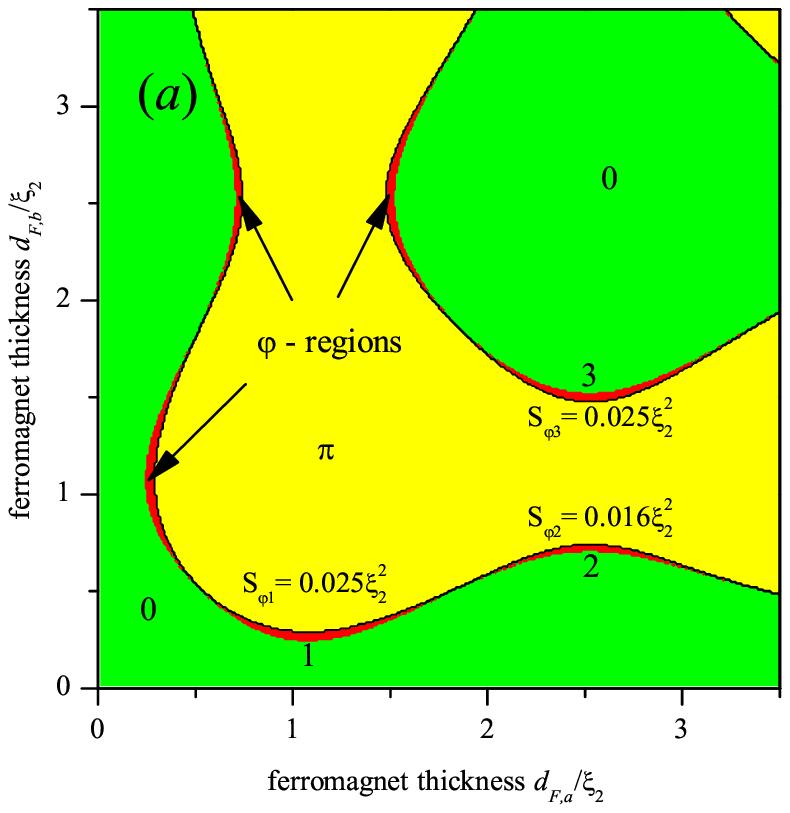}}\hfill
    {\includegraphics{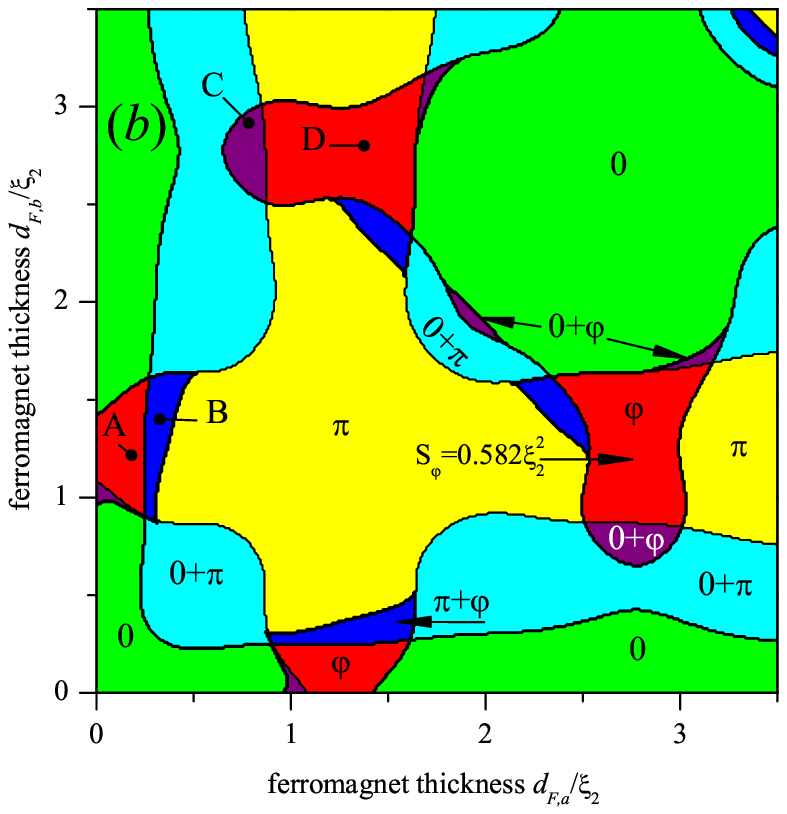}}
  \end{center}
  \caption{(Color online)
    Ground states scheme for the multifacet SFS junction with F-layer thickness periodically changing between $d_{F,a}$ and $d_{F,b}$ plotted on the ($d_{F,a}$, $d_{F,b}$) plane. The values of the ground state phase corresponding to each region are shown, and the areas $S_{\varphi}$ for some of the different $\varphi$ ground state regions are also indicated. In (a) the calculation includes only the first harmonics in every uniform part, i.e., two generated harmonics. The $\varphi$- regions 1,2 and 3 in (a) correspond to the pairs of $d_{F,a}$ and $d_{F,b}$ that are shown by arcs in Fig.~\ref{fig:j(d)}(b).
    In (b) the first 3 harmonics and the corresponding 4 generated harmonics are taken into account, see Eqs.~\eqref{CPR4} and \eqref{J1-4}. The dependence $E_J(\psi)$ for points A,B,C,D is shown in Fig.~\ref{fig:JosEn}.
  }
  \label{fig:GroundStates}
\end{figure*}

In the framework of the clean SFS junction model\cite{Buzdin:1982} we investigate a multifacet junction with $a=b=\lambda _{J}\ll \Lambda _{J}$. For $\lambda _{J}$ we take the value corresponding to the first minimum of $j_{1}$ at $d_F\approx \xi_2$, see Fig.~\ref{fig:j(d)}(a). For clean SFS junctions this value is \cite{Blum:2002:IcOscillations} $\approx 10 \units{kA/cm^2}$, which corresponds to $\lambda _{J}\approx 3 \units{\mu m}$ and which allows to realize $a\approx b$ with reasonable precision. Below we investigate the ground states in a Josephson junction with alternating regions of length $a$, $b$ and F-layer thicknesses $d_{F,a}$ and $d_{F,b}$ varying from 0 to a few $\xi _{2}$.

The ground state corresponds to a local minimum of the energy (\ref{Ej}) with the CPR (\ref{J(ksi)}). The junction has a stable static solution $\psi =0$ (0-phase) if
\begin{equation}
\sum\limits_{n=1}^{N}\left[ n\left\langle j_{n}\right\rangle -\alpha
\sum\limits_{k=1}^{N+1-n}nk\delta j_{n}\delta j_{k}\right] >0,  \label{0-JJ}
\end{equation}
and the solution $\psi =\pi $ ($\pi$-phase) if
\begin{equation}
\sum\limits_{n=1}^{N}\left[ n\left\langle j_{n}\right\rangle -\alpha
\sum\limits_{k=1}^{N+1-n}(-1)^{n+k}nk\delta j_{n}\delta j_{k}\right] >0\;.
\label{pi-JJ}
\end{equation}
For the CPR (\ref{CPR4}) these conditions have the following form: For the 0-phase
\begin{equation}
  J_{1}+2J_{2}+3J_{3}+4J_{4}>0
  , \label{Eq:0-sol}
\end{equation}
and for the $\pi$-phase
\begin{equation}
  J_{1}-2J_{2}+3J_{3}-4J_{4}<0
  . \label{Eq:pi-sol}
\end{equation}
Both solutions $\psi=0$ and $\psi=\pi$ coexist when these conditions are satisfied simultaneously, i.e.
\begin{equation}
  2J_{2}+4J_{4}>\left| J_{1}+3J_{3} \right|
  . \label{Eq:pi+0-sol}
\end{equation}
If both conditions \eqref{Eq:0-sol} and \eqref{Eq:pi-sol} are not satisfied, i.e.
\begin{equation}
  2J_{2}+4J_{4} < -\left| J_{1}+3J_{3} \right|
  , \label{Eq:varphi-sol}
\end{equation}
only the $\varphi$ ground state is possible. The conditions (\ref{Eq:0-sol})--(\ref{Eq:varphi-sol}) coincide with the conditions (\ref{condition2}) if $J_3=J_4=0$.
Generally, the $\varphi$ junction is realized if the Josephson energy $E_J(\psi)$ (\ref{Ej}) has a local minimum, i.e., if the CPR $J(\psi)$ (\ref{CPR4}) crosses $J=0$ from a negative to a positive value at some point $\psi=\varphi\neq 0,\pi $. This is the case if the equation
\begin{equation}
  J(\psi ) \sim 8J_{4}z^{3}+4J_{3}z^{2}+(2J_{2}-4J_{4})z+J_{1}-J_{3}=0
  , \label{Eq:varphi1-sol}
\end{equation}
has at least one real solution $z=\cos \psi $, satisfying the conditions $\left\vert z\right\vert <1$ (which gives $0<\left\vert \psi \right\vert <\pi $), and
\begin{eqnarray}
  \fracp{J}{\psi} &\sim& 32J_{4}z^{4} + 12J_{3}z^{3}+(4J_{2}-8J_{4})z^{2}+
  \nonumber\\
  &+& (J_{1}-9J_{3})z-2J_{2}+4J_{4}>0
  , \label{Eq:varphi2-sol}
\end{eqnarray}
that ensures the local minimum of $E_{J}(\psi)$. Starting from a pair of $d_{F,a},\,d_{F,b}$ we calculate $J_{1\ldots 4}$. Then from Eqs.~(\ref{Eq:0-sol})--(\ref{Eq:varphi2-sol}) the possible ground states
are identified. The resulting phase diagram is shown in Fig.~\ref{fig:GroundStates}, where different ground states for each pair of $d_{F,a},\,d_{F,b}$ are shown by different colors.

Figure \ref{fig:GroundStates}(a) shows the results obtained if only the first intrinsic harmonic is taken into account. Here, the areas of 0 and $\pi$ ground state phase are separated by slim regions of $\varphi$ phase. It is clear that the phase diagram is symmetric with respect to the line $d_{F,a}=d_{F,b}$. Therefore, below, without loosing generality, we focus on the case $d_{F,a}>d_{F,b}$. In the chosen interval of thicknesses there are 3 areas of $\varphi$ phase, marked as 1,2 and 3. In these areas $j_{1}(d_{F,a})\approx -j_{1}(d_{F,b})$, which corresponds to the pairs of $d_{F,a}$ and $d_{F,b}$ shown in Fig.~\ref{fig:j(d)}(b) as 1,2 and 3. If $d_{F,a}$ and $d_{F,b}$ are not well controllable, the area $S_{\varphi}$ of $\varphi$ regions is proportional to the probability of the $\varphi$ junction realization. For all three different cases shown in Fig.~\ref{fig:GroundStates}(a) $S_{\varphi}$ is rather small ($\le 0.025\xi_2^2$). Hence, one has to control $d_{F,a}$ and $d_{F,b}$ extremely precisely in order to realize a $\varphi$ junction. In the case of a ``dirty'' SFS junction, when $j_{1}(d_F)$ decays exponentially, the area $S_{\varphi}$ is even smaller, and the probability to fabricate a $\varphi $-junction is vanishing.

However, if we take into account a non-sinusoidal CPR for the ``clean'' SFS junction, the areas of the $\varphi$ ground state become much larger, as can be seen in Fig.~\ref{fig:GroundStates}(b). This is a consequence of the fact that the intrinsic second harmonic $j_{2}(d_F)<0$ in the corresponding regions, which efficiently helps to make the absolute value of the generated second harmonic large enough.

In particular, the areas 2 and 3 from Fig.~\ref{fig:GroundStates}(a) merge
%
%
and form a compact $\varphi$ ground state region with area $S_\varphi\approx 0.6\xi_2 ^2$ around $d_{F,a}\sim 2.8\xi_2$ and $d_{F,b}\sim 1.4\xi_2$. This region seems to be very well suited for the experimental realization of a $\varphi$ junction, as it does not demand to produce an extremely thin F-layer and at the same time allows for reasonably large tolerances in sample fabrication. Taking the ferromagnetic coherence length $\xi_{2}\sim 1.2\units{nm}$ from experiments\cite{RobinsonPRB76,Blum:2002:IcOscillations},
the linear size of the $\varphi$ region in Fig.\ref{fig:GroundStates}(b) is $\Delta d_{F,a}\sim\Delta d_{F,b}\sim 0.5\ldots 1 \units{nm}$.
Modern technology allows the control of $d_F$ with such precision \cite{Weides:SIFS-Tech,Ryazanov0-piJJ,Born:2006:SIFS-Ni3Al}.

\begin{figure}[!htb]
  \begin{center}
    \includegraphics{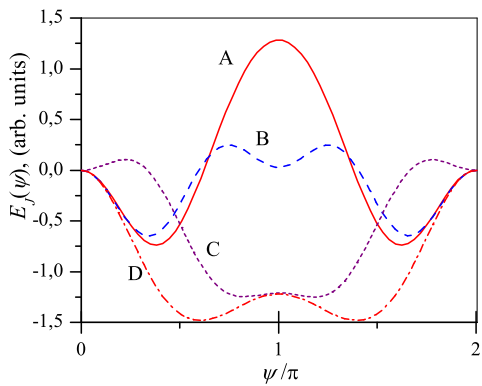}
  \end{center}
  \caption{(Color online)
    Josephson energy $E_J(\psi)$ of multifacet SFS junctions for different combinations of F-layer thicknesses $d_{F,a}$ and $d_{F,b}$ as indicated by points A,B,C, and D in Fig.~\ref{fig:GroundStates}(b).
    A and D correspond to the $\varphi$ ground state; B
    corresponds to two stable states $\pi$ and $\varphi$; and C corresponds to two stable states $0$ and $\varphi$.
  }
  \label{fig:JosEn}
\end{figure}

Finally we note that the structure considered here may have simultaneously two different stable static solutions: 0 and $\pi$, 0 and $\varphi$, or $\pi$ and $\varphi $, as can be seen in Fig.~\ref{fig:GroundStates}(b). The corresponding $E_{J}(\psi)$ curves
are presented in Fig.~\ref{fig:JosEn}, where one can see several local minima of the Josephson energy as a function of the phase. Probably, taking into account the next harmonics,
it is also possible to find two different $\varphi $ ground states. However, we expect the corresponding regions within the $d_{F,a}, d_{F,b}$ plane to be so small that it will be quite hard to fabricate such a structure.


\section{Conclusion}
\label{sec:Con}

We propose a method to realize a $\varphi$ junction, i.e.~with the Josephson phase $\varphi\neq 0$ or $\pi$ in the ground state, based on SFS junctions.
In a uniform SFS Josephson junction near a 0-$\pi$ transition (with appropriate F-layer thickness $d_F$) the first harmonic $j_1\to 0$, while the second harmonic $j_2$ dominates, however usually with positive sign ($j_2>0$), which excludes the formation of a $\varphi$ ground state.
Instead, the main idea of the method used here to make a $\varphi$ junction is the following: We choose a thickness $d_F$ where the second harmonic is large and negative. To cancel the first harmonic we use a periodic step-like modulation of the F-layer thickness between $d_{F,a}$ and $d_{F,b}$.  Here, $d_{F,a}$ and $d_{F,b}$ are chosen such that for both of them the second harmonic is negative, while the first harmonic is $j_1(d_{F,a})\approx-j_1(d_{F,b})$. Periodic modulation not only cancels the first harmonic but also generates an additional negative second harmonic for the average phase. This effect can be made stronger if one works with a ``clean'' ferromagnetic barrier and uses alternating regions of equal length $a \approx b \sim \lambda_{J}$.

Different mechanisms, leading to a significantly non-sinusoidal current-phase relation of SFS junctions are analyzed. The CPR mostly different from the sinusoidal one is obtained for a ``clean'' SFS junction at low temperature. In this case, there are reasonably large regions of thicknesses $d_{F,a}$ and $d_{F,b}$ (in comparison with the description taking into account only the first harmonic)
where the multifacet Josephson junction has a $\varphi $-ground state.
Moreover, for some values of $d_{F,a}$ and $d_{F,b}$ such structures may have two different ground states (two local
minima of the Josephson energy as a function of the phase): 0 and $\pi $, 0 and $\varphi $, or $\pi$ and $\varphi $. Our analysis gives some practical recommendations for the fabrication of SFS junctions with arbitrary phase shifts $\varphi $ in the ground state.
\bigskip

\begin{acknowledgments}

We gratefully acknowledge M.~Yu.~Kupriyanov and V.~V.~Ryazanov for helpful discussions. This work was supported by the Russian Foundation for Basic Research (Grant 09-02-12176-ofi-m), by the German-Israeli Foundation (Grant G-967-126.14/2007) and by the Deutsche Forschungsgemeinschaft (DFG) via the SFB/TRR 21.

\end{acknowledgments}

\bibliography{pi,SFS2}

\begin{thebibliography}{37}
\expandafter\ifx\csname natexlab\endcsname\relax\def\natexlab#1{#1}\fi
\expandafter\ifx\csname bibnamefont\endcsname\relax
  \def\bibnamefont#1{#1}\fi
\expandafter\ifx\csname bibfnamefont\endcsname\relax
  \def\bibfnamefont#1{#1}\fi
\expandafter\ifx\csname citenamefont\endcsname\relax
  \def\citenamefont#1{#1}\fi
\expandafter\ifx\csname url\endcsname\relax
  \def\url#1{\texttt{#1}}\fi
\expandafter\ifx\csname urlprefix\endcsname\relax\def\urlprefix{URL }\fi
\providecommand{\bibinfo}[2]{#2}
\providecommand{\eprint}[2][]{\url{#2}}

\bibitem[{\citenamefont{Buzdin}(2005{\natexlab{a}})}]{Buzdin:2005:Review:SF}
\bibinfo{author}{\bibfnamefont{A.~I.} \bibnamefont{Buzdin}},
  \bibinfo{journal}{Rev. Mod. Phys.} \textbf{\bibinfo{volume}{77}},
  \bibinfo{eid}{935} (\bibinfo{year}{2005}{\natexlab{a}}).

\bibitem[{\citenamefont{Golubov et~al.}(2004)\citenamefont{Golubov, Kupriyanov,
  and Il'ichev}}]{KuprReiew}
\bibinfo{author}{\bibfnamefont{A.}~\bibnamefont{Golubov}},
  \bibinfo{author}{\bibfnamefont{M.}~\bibnamefont{Kupriyanov}},
  \bibnamefont{and} \bibinfo{author}{\bibfnamefont{E.}~\bibnamefont{Il'ichev}},
  \bibinfo{journal}{Rev. Mod. Phys.} \textbf{\bibinfo{volume}{76}},
  \bibinfo{pages}{411} (\bibinfo{year}{2004}).

\bibitem[{\citenamefont{Ortlepp et~al.}(2006)\citenamefont{Ortlepp, Ariando,
  Mielke, Verwijs, Foo, Rogalla, Uhlmann, and
  Hilgenkamp}}]{Ortlepp:2006:RSFQ-0-pi}
\bibinfo{author}{\bibfnamefont{T.}~\bibnamefont{Ortlepp}},
  \bibinfo{author}{\bibnamefont{Ariando}},
  \bibinfo{author}{\bibfnamefont{O.}~\bibnamefont{Mielke}},
  \bibinfo{author}{\bibfnamefont{C.~J.~M.} \bibnamefont{Verwijs}},
  \bibinfo{author}{\bibfnamefont{K.~F.~K.} \bibnamefont{Foo}},
  \bibinfo{author}{\bibfnamefont{H.}~\bibnamefont{Rogalla}},
  \bibinfo{author}{\bibfnamefont{F.~H.} \bibnamefont{Uhlmann}},
  \bibnamefont{and}
  \bibinfo{author}{\bibfnamefont{H.}~\bibnamefont{Hilgenkamp}},
  \bibinfo{journal}{Science} \textbf{\bibinfo{volume}{312}},
  \bibinfo{pages}{1495} (\bibinfo{year}{2006}).

\bibitem[{\citenamefont{Ioffe et~al.}(1999)\citenamefont{Ioffe, Geshkenbein,
  Feigel'man, Fauche\`ere, and Blatter}}]{Ioffe:1999:sds-waveQubit}
\bibinfo{author}{\bibfnamefont{L.~B.} \bibnamefont{Ioffe}},
  \bibinfo{author}{\bibfnamefont{V.~B.} \bibnamefont{Geshkenbein}},
  \bibinfo{author}{\bibfnamefont{M.~V.} \bibnamefont{Feigel'man}},
  \bibinfo{author}{\bibfnamefont{A.~L.} \bibnamefont{Fauche\`ere}},
  \bibnamefont{and} \bibinfo{author}{\bibfnamefont{G.}~\bibnamefont{Blatter}},
  \bibinfo{journal}{Nature (London)} \textbf{\bibinfo{volume}{398}},
  \bibinfo{pages}{679} (\bibinfo{year}{1999}).

\bibitem[{\citenamefont{Klenov et~al.}(2008)\citenamefont{Klenov, Kornev,
  Vedyayev, Ryzhanova, Pugach, and Rumyantseva}}]{OurQubit}
\bibinfo{author}{\bibfnamefont{N.}~\bibnamefont{Klenov}},
  \bibinfo{author}{\bibfnamefont{V.}~\bibnamefont{Kornev}},
  \bibinfo{author}{\bibfnamefont{A.}~\bibnamefont{Vedyayev}},
  \bibinfo{author}{\bibfnamefont{N.}~\bibnamefont{Ryzhanova}},
  \bibinfo{author}{\bibfnamefont{N.}~\bibnamefont{Pugach}}, \bibnamefont{and}
  \bibinfo{author}{\bibfnamefont{T.}~\bibnamefont{Rumyantseva}},
  \bibinfo{journal}{J. Phys. Conf. Ser.} \textbf{\bibinfo{volume}{97}},
  \bibinfo{pages}{012037} (\bibinfo{year}{2008}).

\bibitem[{\citenamefont{Yamashita et~al.}(2005)\citenamefont{Yamashita,
  Tanikawa, Takahashi, and Maekawa}}]{Yamashita:2005:pi-qubit:SFS+SIS}
\bibinfo{author}{\bibfnamefont{T.}~\bibnamefont{Yamashita}},
  \bibinfo{author}{\bibfnamefont{K.}~\bibnamefont{Tanikawa}},
  \bibinfo{author}{\bibfnamefont{S.}~\bibnamefont{Takahashi}},
  \bibnamefont{and} \bibinfo{author}{\bibfnamefont{S.}~\bibnamefont{Maekawa}},
  \bibinfo{journal}{Phys. Rev. Lett.} \textbf{\bibinfo{volume}{95}},
  \bibinfo{eid}{097001} (\bibinfo{year}{2005}).

\bibitem[{\citenamefont{Yamashita et~al.}(2006)\citenamefont{Yamashita,
  Takahashi, and Maekawa}}]{Yamashita:2006:pi-qubit:3JJ}
\bibinfo{author}{\bibfnamefont{T.}~\bibnamefont{Yamashita}},
  \bibinfo{author}{\bibfnamefont{S.}~\bibnamefont{Takahashi}},
  \bibnamefont{and} \bibinfo{author}{\bibfnamefont{S.}~\bibnamefont{Maekawa}},
  \bibinfo{journal}{Appl. Phys. Lett.} \textbf{\bibinfo{volume}{88}},
  \bibinfo{eid}{132501} (\bibinfo{year}{2006}).

\bibitem[{\citenamefont{Ustinov and
  Kaplunenko}(2003)}]{Ustinov:2003:RSFQ+pi-shifters}
\bibinfo{author}{\bibfnamefont{A.~V.} \bibnamefont{Ustinov}} \bibnamefont{and}
  \bibinfo{author}{\bibfnamefont{V.~K.} \bibnamefont{Kaplunenko}},
  \bibinfo{journal}{J. Appl. Phys.} \textbf{\bibinfo{volume}{94}},
  \bibinfo{pages}{5405} (\bibinfo{year}{2003}).

\bibitem[{\citenamefont{Feofanov et~al.}()\citenamefont{Feofanov, Oboznov,
  Bolginov, Lisenfeld, Poletto, Ryazanov, Rossolenko, Khabipov, Balashov, Zorin
  et~al.}}]{RyazanovPiCirc2009}
\bibinfo{author}{\bibfnamefont{A.~K.} \bibnamefont{Feofanov}},
  \bibinfo{author}{\bibfnamefont{V.~A.} \bibnamefont{Oboznov}},
  \bibinfo{author}{\bibfnamefont{V.}~\bibnamefont{Bolginov}},
  \bibinfo{author}{\bibfnamefont{J.}~\bibnamefont{Lisenfeld}},
  \bibinfo{author}{\bibfnamefont{S.}~\bibnamefont{Poletto}},
  \bibinfo{author}{\bibfnamefont{V.~V.} \bibnamefont{Ryazanov}},
  \bibinfo{author}{\bibfnamefont{A.~N.} \bibnamefont{Rossolenko}},
  \bibinfo{author}{\bibfnamefont{M.}~\bibnamefont{Khabipov}},
  \bibinfo{author}{\bibfnamefont{D.}~\bibnamefont{Balashov}},
  \bibinfo{author}{\bibfnamefont{A.~B.} \bibnamefont{Zorin}},
  \bibnamefont{et~al.}, \bibinfo{note}{submitted to Nature Phys.}

\bibitem[{\citenamefont{Buzdin and Koshelev}(2003)}]{BuzdinKoshelev}
\bibinfo{author}{\bibfnamefont{A.}~\bibnamefont{Buzdin}} \bibnamefont{and}
  \bibinfo{author}{\bibfnamefont{A.~E.} \bibnamefont{Koshelev}},
  \bibinfo{journal}{Phys. Rev. B} \textbf{\bibinfo{volume}{67}},
  \bibinfo{eid}{220504(R)} (\bibinfo{year}{2003}).

\bibitem[{\citenamefont{Goldobin et~al.}(2007)\citenamefont{Goldobin, Koelle,
  Kleiner, and Buzdin}}]{Goldobin2harm}
\bibinfo{author}{\bibfnamefont{E.}~\bibnamefont{Goldobin}},
  \bibinfo{author}{\bibfnamefont{D.}~\bibnamefont{Koelle}},
  \bibinfo{author}{\bibfnamefont{R.}~\bibnamefont{Kleiner}}, \bibnamefont{and}
  \bibinfo{author}{\bibfnamefont{A.}~\bibnamefont{Buzdin}},
  \bibinfo{journal}{Phys. Rev. B} \textbf{\bibinfo{volume}{76}},
  \bibinfo{pages}{224523} (\bibinfo{year}{2007}).

\bibitem[{\citenamefont{Ryazanov et~al.}(2008)\citenamefont{Ryazanov, Oboznov,
  Bolginov, and Rossolenko}}]{Ryazanov2harmNovgorod}
\bibinfo{author}{\bibfnamefont{V.~V.} \bibnamefont{Ryazanov}},
  \bibinfo{author}{\bibfnamefont{V.~A.} \bibnamefont{Oboznov}},
  \bibinfo{author}{\bibfnamefont{V.}~\bibnamefont{Bolginov}}, \bibnamefont{and}
  \bibinfo{author}{\bibfnamefont{A.}~\bibnamefont{Rossolenko}}, in
  \emph{\bibinfo{booktitle}{Proceedings of XII International Symphosium
  ``Nanophysics and Nanoelectronics''}} (\bibinfo{publisher}{IFM RAS},
  \bibinfo{address}{Nizhniy Novgorod}, \bibinfo{year}{2008}),
  vol.~\bibinfo{volume}{1}, p.~\bibinfo{pages}{42}, \bibinfo{note}{(in
  Russian)}.

\bibitem[{\citenamefont{Sellier et~al.}(2004)\citenamefont{Sellier, Baraduc,
  Lefloch, and Calemczuk}}]{Sellier:2004:SFS:HalfIntShapiro}
\bibinfo{author}{\bibfnamefont{H.}~\bibnamefont{Sellier}},
  \bibinfo{author}{\bibfnamefont{C.}~\bibnamefont{Baraduc}},
  \bibinfo{author}{\bibfnamefont{F.}~\bibnamefont{Lefloch}}, \bibnamefont{and}
  \bibinfo{author}{\bibfnamefont{R.}~\bibnamefont{Calemczuk}},
  \bibinfo{journal}{Phys. Rev. Lett.} \textbf{\bibinfo{volume}{92}},
  \bibinfo{eid}{257005} (\bibinfo{year}{2004}).

\bibitem[{\citenamefont{Buzdin et~al.}(1982)\citenamefont{Buzdin, Bulaevskii,
  and Panyukov}}]{Buzdin:1982}
\bibinfo{author}{\bibfnamefont{A.~I.} \bibnamefont{Buzdin}},
  \bibinfo{author}{\bibfnamefont{L.}~\bibnamefont{Bulaevskii}},
  \bibnamefont{and} \bibinfo{author}{\bibfnamefont{S.}~\bibnamefont{Panyukov}},
  \bibinfo{journal}{JETP Lett.} \textbf{\bibinfo{volume}{35}},
  \bibinfo{pages}{178} (\bibinfo{year}{1982}), \bibinfo{note}{pis'ma v ZhETF
  \textbf{35} p147 (1982)}.

\bibitem[{\citenamefont{Konschelle et~al.}(2008)\citenamefont{Konschelle,
  Cayssol, and Buzdin}}]{Buzdin2-3D}
\bibinfo{author}{\bibfnamefont{F.}~\bibnamefont{Konschelle}},
  \bibinfo{author}{\bibfnamefont{J.}~\bibnamefont{Cayssol}}, \bibnamefont{and}
  \bibinfo{author}{\bibfnamefont{A.~I.} \bibnamefont{Buzdin}},
  \bibinfo{journal}{Phys. Rev. B} \textbf{\bibinfo{volume}{78}},
  \bibinfo{pages}{134505} (\bibinfo{year}{2008}).

\bibitem[{\citenamefont{Buzdin}(2005{\natexlab{b}})}]{Buzdin:2005:0-pi-trans}
\bibinfo{author}{\bibfnamefont{A.}~\bibnamefont{Buzdin}},
  \bibinfo{journal}{Phys. Rev. B} \textbf{\bibinfo{volume}{72}},
  \bibinfo{eid}{100501(R)} (\bibinfo{year}{2005}{\natexlab{b}}).

\bibitem[{\citenamefont{Vedyayev et~al.}(2006)\citenamefont{Vedyayev,
  Ryzhanova, and Pugach}}]{OurSFS}
\bibinfo{author}{\bibfnamefont{A.}~\bibnamefont{Vedyayev}},
  \bibinfo{author}{\bibfnamefont{N.}~\bibnamefont{Ryzhanova}},
  \bibnamefont{and} \bibinfo{author}{\bibfnamefont{N.}~\bibnamefont{Pugach}},
  \bibinfo{journal}{J. Magn. Magn. Mat.} \textbf{\bibinfo{volume}{305}},
  \bibinfo{pages}{53} (\bibinfo{year}{2006}).

\bibitem[{\citenamefont{Golubov and Kupriyanov}(2005)}]{KuprCPR}
\bibinfo{author}{\bibfnamefont{A.}~\bibnamefont{Golubov}} \bibnamefont{and}
  \bibinfo{author}{\bibfnamefont{M.}~\bibnamefont{Kupriyanov}},
  \bibinfo{journal}{Pis'ma v ZhETF} \textbf{\bibinfo{volume}{81}},
  \bibinfo{pages}{419} (\bibinfo{year}{2005}), \bibinfo{note}{{J}ETP Letters
  \textbf{81}, 335 (2005)}.

\bibitem[{\citenamefont{Golubov et~al.}(2002)\citenamefont{Golubov, Kupriyanov,
  and Fominov}}]{KuprFominovCPR}
\bibinfo{author}{\bibfnamefont{A.}~\bibnamefont{Golubov}},
  \bibinfo{author}{\bibfnamefont{M.}~\bibnamefont{Kupriyanov}},
  \bibnamefont{and} \bibinfo{author}{\bibfnamefont{Y.~V.}
  \bibnamefont{Fominov}}, \bibinfo{journal}{Pis'ma v ZhETF}
  \textbf{\bibinfo{volume}{75}}, \bibinfo{pages}{709} (\bibinfo{year}{2002}),
  \bibinfo{note}{{J}ETP Letters \textbf{75}, 588 (2002)}.

\bibitem[{\citenamefont{Mints}(1998)}]{Mints2harm}
\bibinfo{author}{\bibfnamefont{R.~G.} \bibnamefont{Mints}},
  \bibinfo{journal}{Phys. Rev. B} \textbf{\bibinfo{volume}{57}},
  \bibinfo{pages}{R3221} (\bibinfo{year}{1998}).

\bibitem[{\citenamefont{Weides et~al.}(2006{\natexlab{a}})\citenamefont{Weides,
  Kemmler, Kohlstedt, Waser, Koelle, Kleiner, and Goldobin}}]{Weides:0-piLJJ}
\bibinfo{author}{\bibfnamefont{M.}~\bibnamefont{Weides}},
  \bibinfo{author}{\bibfnamefont{M.}~\bibnamefont{Kemmler}},
  \bibinfo{author}{\bibfnamefont{H.}~\bibnamefont{Kohlstedt}},
  \bibinfo{author}{\bibfnamefont{R.}~\bibnamefont{Waser}},
  \bibinfo{author}{\bibfnamefont{D.}~\bibnamefont{Koelle}},
  \bibinfo{author}{\bibfnamefont{R.}~\bibnamefont{Kleiner}}, \bibnamefont{and}
  \bibinfo{author}{\bibfnamefont{E.}~\bibnamefont{Goldobin}},
  \bibinfo{journal}{Phys. Rev. Lett.} \textbf{\bibinfo{volume}{97}},
  \bibinfo{eid}{247001} (\bibinfo{year}{2006}{\natexlab{a}}).

\bibitem[{\citenamefont{Smilde et~al.}(2002)\citenamefont{Smilde, Ariando,
  Blank, Gerritsma, Hilgenkamp, and Rogalla}}]{Smilde:ZigzagPRL}
\bibinfo{author}{\bibfnamefont{H.-J.~H.} \bibnamefont{Smilde}},
  \bibinfo{author}{\bibnamefont{Ariando}},
  \bibinfo{author}{\bibfnamefont{D.~H.~A.} \bibnamefont{Blank}},
  \bibinfo{author}{\bibfnamefont{G.~J.} \bibnamefont{Gerritsma}},
  \bibinfo{author}{\bibfnamefont{H.}~\bibnamefont{Hilgenkamp}},
  \bibnamefont{and} \bibinfo{author}{\bibfnamefont{H.}~\bibnamefont{Rogalla}},
  \bibinfo{journal}{Phys. Rev. Lett.} \textbf{\bibinfo{volume}{88}},
  \bibinfo{pages}{057004} (\bibinfo{year}{2002}).

\bibitem[{\citenamefont{Mints and
  Papiashvili}(2000)}]{Mints:2000:SelfGenFlux@GB}
\bibinfo{author}{\bibfnamefont{R.~G.} \bibnamefont{Mints}} \bibnamefont{and}
  \bibinfo{author}{\bibfnamefont{I.}~\bibnamefont{Papiashvili}},
  \bibinfo{journal}{Phys. Rev. B} \textbf{\bibinfo{volume}{62}},
  \bibinfo{pages}{15214} (\bibinfo{year}{2000}).

\bibitem[{\citenamefont{Mints and
  Papiashvili}(2001)}]{Mints:2001:FracVortices@GB}
\bibinfo{author}{\bibfnamefont{R.~G.} \bibnamefont{Mints}} \bibnamefont{and}
  \bibinfo{author}{\bibfnamefont{I.}~\bibnamefont{Papiashvili}},
  \bibinfo{journal}{Phys. Rev. B} \textbf{\bibinfo{volume}{64}},
  \bibinfo{pages}{134501} (\bibinfo{year}{2001}).

\bibitem[{\citenamefont{Landau and Lifshits}(1960)}]{LLmechanics}
\bibinfo{author}{\bibfnamefont{L.}~\bibnamefont{Landau}} \bibnamefont{and}
  \bibinfo{author}{\bibfnamefont{E.}~\bibnamefont{Lifshits}},
  \emph{\bibinfo{title}{Mechanics}} (\bibinfo{publisher}{Pergamon Press},
  \bibinfo{address}{Oxford}, \bibinfo{year}{1960}).

\bibitem[{\citenamefont{Kupriyanov et~al.}(2008)\citenamefont{Kupriyanov,
  Pugach, Khapaev, Vedyayev, Goldobin, Koelle, and Kleiner}}]{ourJETPL}
\bibinfo{author}{\bibfnamefont{M.}~\bibnamefont{Kupriyanov}},
  \bibinfo{author}{\bibfnamefont{N.}~\bibnamefont{Pugach}},
  \bibinfo{author}{\bibfnamefont{M.}~\bibnamefont{Khapaev}},
  \bibinfo{author}{\bibfnamefont{A.}~\bibnamefont{Vedyayev}},
  \bibinfo{author}{\bibfnamefont{E.}~\bibnamefont{Goldobin}},
  \bibinfo{author}{\bibfnamefont{D.}~\bibnamefont{Koelle}}, \bibnamefont{and}
  \bibinfo{author}{\bibfnamefont{R.}~\bibnamefont{Kleiner}},
  \bibinfo{journal}{Pis'ma v ZhETF} \textbf{\bibinfo{volume}{88}},
  \bibinfo{pages}{50} (\bibinfo{year}{2008}), \bibinfo{note}{{J}ETP Lett.
  \textbf{88}, 45 (2008)}.

\bibitem[{\citenamefont{Vedyayev et~al.}(2005)\citenamefont{Vedyayev, Lacroix,
  Pugach, and Ryzhanova}}]{OurEuLett}
\bibinfo{author}{\bibfnamefont{A.}~\bibnamefont{Vedyayev}},
  \bibinfo{author}{\bibfnamefont{C.}~\bibnamefont{Lacroix}},
  \bibinfo{author}{\bibfnamefont{N.}~\bibnamefont{Pugach}}, \bibnamefont{and}
  \bibinfo{author}{\bibfnamefont{N.}~\bibnamefont{Ryzhanova}},
  \bibinfo{journal}{Europhys. Lett.} \textbf{\bibinfo{volume}{71}},
  \bibinfo{pages}{679} (\bibinfo{year}{2005}).

\bibitem[{\citenamefont{Oboznov et~al.}(2006)\citenamefont{Oboznov, Bolginov,
  Feofanov, Ryazanov, and Buzdin}}]{Oboznov:2006:SFS-Ic(dF)}
\bibinfo{author}{\bibfnamefont{V.~A.} \bibnamefont{Oboznov}},
  \bibinfo{author}{\bibfnamefont{V.~V.} \bibnamefont{Bolginov}},
  \bibinfo{author}{\bibfnamefont{A.~K.} \bibnamefont{Feofanov}},
  \bibinfo{author}{\bibfnamefont{V.~V.} \bibnamefont{Ryazanov}},
  \bibnamefont{and} \bibinfo{author}{\bibfnamefont{A.~I.}
  \bibnamefont{Buzdin}}, \bibinfo{journal}{Phys. Rev. Lett.}
  \textbf{\bibinfo{volume}{96}}, \bibinfo{eid}{197003} (\bibinfo{year}{2006}).

\bibitem[{\citenamefont{Bergeret et~al.}(2001)\citenamefont{Bergeret, Volkov,
  and Efetov}}]{BergeretVolkovEfetovPRB2001}
\bibinfo{author}{\bibfnamefont{F.~S.} \bibnamefont{Bergeret}},
  \bibinfo{author}{\bibfnamefont{A.~F.} \bibnamefont{Volkov}},
  \bibnamefont{and} \bibinfo{author}{\bibfnamefont{K.~B.}
  \bibnamefont{Efetov}}, \bibinfo{journal}{Phys. Rev. B}
  \textbf{\bibinfo{volume}{64}}, \bibinfo{pages}{134506}
  (\bibinfo{year}{2001}).

\bibitem[{\citenamefont{Robinson et~al.}(2006)\citenamefont{Robinson, Piano,
  Burnell, Bell, and Blamire}}]{RobinsonPRL97}
\bibinfo{author}{\bibfnamefont{J.~W.~A.} \bibnamefont{Robinson}},
  \bibinfo{author}{\bibfnamefont{S.}~\bibnamefont{Piano}},
  \bibinfo{author}{\bibfnamefont{G.}~\bibnamefont{Burnell}},
  \bibinfo{author}{\bibfnamefont{C.}~\bibnamefont{Bell}}, \bibnamefont{and}
  \bibinfo{author}{\bibfnamefont{M.~G.} \bibnamefont{Blamire}},
  \bibinfo{journal}{Phys. Rev. Lett.} \textbf{\bibinfo{volume}{97}},
  \bibinfo{pages}{177003} (\bibinfo{year}{2006}).

\bibitem[{\citenamefont{Robinson et~al.}(2007)\citenamefont{Robinson, Piano,
  Burnell, Bell, and Blamire}}]{RobinsonPRB76}
\bibinfo{author}{\bibfnamefont{J.~W.~A.} \bibnamefont{Robinson}},
  \bibinfo{author}{\bibfnamefont{S.}~\bibnamefont{Piano}},
  \bibinfo{author}{\bibfnamefont{G.}~\bibnamefont{Burnell}},
  \bibinfo{author}{\bibfnamefont{C.}~\bibnamefont{Bell}}, \bibnamefont{and}
  \bibinfo{author}{\bibfnamefont{M.~G.} \bibnamefont{Blamire}},
  \bibinfo{journal}{Phys. Rev. B} \textbf{\bibinfo{volume}{76}},
  \bibinfo{pages}{094522} (\bibinfo{year}{2007}).

\bibitem[{\citenamefont{Blum et~al.}(2002)\citenamefont{Blum, Tsukernik,
  Karpovski, and Palevski}}]{Blum:2002:IcOscillations}
\bibinfo{author}{\bibfnamefont{Y.}~\bibnamefont{Blum}},
  \bibinfo{author}{\bibfnamefont{A.}~\bibnamefont{Tsukernik}},
  \bibinfo{author}{\bibfnamefont{M.}~\bibnamefont{Karpovski}},
  \bibnamefont{and} \bibinfo{author}{\bibfnamefont{A.}~\bibnamefont{Palevski}},
  \bibinfo{journal}{Phys. Rev. Lett.} \textbf{\bibinfo{volume}{89}},
  \bibinfo{pages}{187004} (\bibinfo{year}{2002}).

\bibitem[{\citenamefont{Bannykh et~al.}(2009)\citenamefont{Bannykh, Pfeiffer,
  Stolyarov, Batov, Ryazanov, and Weides}}]{WeidesRyazanov2009}
\bibinfo{author}{\bibfnamefont{A.~A.} \bibnamefont{Bannykh}},
  \bibinfo{author}{\bibfnamefont{J.}~\bibnamefont{Pfeiffer}},
  \bibinfo{author}{\bibfnamefont{V.~S.} \bibnamefont{Stolyarov}},
  \bibinfo{author}{\bibfnamefont{I.~E.} \bibnamefont{Batov}},
  \bibinfo{author}{\bibfnamefont{V.~V.} \bibnamefont{Ryazanov}},
  \bibnamefont{and} \bibinfo{author}{\bibfnamefont{M.}~\bibnamefont{Weides}},
  \bibinfo{journal}{Phys. Rev. B} \textbf{\bibinfo{volume}{79}},
  \bibinfo{pages}{054501} (\bibinfo{year}{2009}).

\bibitem[{\citenamefont{Radovi\'{c} et~al.}(2001)\citenamefont{Radovi\'{c},
  Dobrosavljevi\'{c}-Gruji\'{c}, and Vuji\v{c}i\'{c}}}]{Radovic:2001:0&pi-JJ}
\bibinfo{author}{\bibfnamefont{Z.}~\bibnamefont{Radovi\'{c}}},
  \bibinfo{author}{\bibfnamefont{L.}~\bibnamefont{Dobrosavljevi\'{c}-Gruji\'{c%
}}}, \bibnamefont{and}
  \bibinfo{author}{\bibfnamefont{B.}~\bibnamefont{Vuji\v{c}i\'{c}}},
  \bibinfo{journal}{Phys. Rev. B} \textbf{\bibinfo{volume}{63}},
  \bibinfo{pages}{214512} (\bibinfo{year}{2001}).

\bibitem[{\citenamefont{Weides et~al.}(2006{\natexlab{b}})\citenamefont{Weides,
  Tillmann, and Kohlstedt}}]{Weides:SIFS-Tech}
\bibinfo{author}{\bibfnamefont{M.}~\bibnamefont{Weides}},
  \bibinfo{author}{\bibfnamefont{K.}~\bibnamefont{Tillmann}}, \bibnamefont{and}
  \bibinfo{author}{\bibfnamefont{H.}~\bibnamefont{Kohlstedt}},
  \bibinfo{journal}{Physica C} \textbf{\bibinfo{volume}{437--438}},
  \bibinfo{pages}{349} (\bibinfo{year}{2006}{\natexlab{b}}).

\bibitem[{\citenamefont{Frolov et~al.}(2006)\citenamefont{Frolov,
  Van~Harlingen, Bolginov, Oboznov, and Ryazanov}}]{Ryazanov0-piJJ}
\bibinfo{author}{\bibfnamefont{S.~M.} \bibnamefont{Frolov}},
  \bibinfo{author}{\bibfnamefont{D.~J.} \bibnamefont{Van~Harlingen}},
  \bibinfo{author}{\bibfnamefont{V.~V.} \bibnamefont{Bolginov}},
  \bibinfo{author}{\bibfnamefont{V.~A.} \bibnamefont{Oboznov}},
  \bibnamefont{and} \bibinfo{author}{\bibfnamefont{V.~V.}
  \bibnamefont{Ryazanov}}, \bibinfo{journal}{Phys. Rev. B}
  \textbf{\bibinfo{volume}{74}}, \bibinfo{pages}{020503(R)}
  (\bibinfo{year}{2006}).

\bibitem[{\citenamefont{Born et~al.}(2006)\citenamefont{Born, Siegel, Hollmann,
  Braak, Golubov, Gusakova, and Kupriyanov}}]{Born:2006:SIFS-Ni3Al}
\bibinfo{author}{\bibfnamefont{F.}~\bibnamefont{Born}},
  \bibinfo{author}{\bibfnamefont{M.}~\bibnamefont{Siegel}},
  \bibinfo{author}{\bibfnamefont{E.~K.} \bibnamefont{Hollmann}},
  \bibinfo{author}{\bibfnamefont{H.}~\bibnamefont{Braak}},
  \bibinfo{author}{\bibfnamefont{A.~A.} \bibnamefont{Golubov}},
  \bibinfo{author}{\bibfnamefont{D.~Y.} \bibnamefont{Gusakova}},
  \bibnamefont{and} \bibinfo{author}{\bibfnamefont{M.~Y.}
  \bibnamefont{Kupriyanov}}, \bibinfo{journal}{Phys. Rev. B}
  \textbf{\bibinfo{volume}{74}}, \bibinfo{eid}{140501(R)}
  (\bibinfo{year}{2006}).

\end{thebibliography}

 \end{document}